\documentclass[a4paper]{jpconf}
\usepackage{graphicx}
\begin{document}
\title{Forward jet-like event spin-dependent properties in polarized p+p collisions at $\sqrt{s}$= 200 GeV}

\author{N Poljak for the STAR collaboration}


\ead{nikola@rcf.rhic.bnl.gov}

\begin{abstract}
The STAR collaboration has reported precision measurements on the transverse single spin asymmetries for the production of forward $\pi^0$ mesons from polarized proton collisions at $\sqrt{s} =\,$200 GeV. To disentangle the contributions to forward asymmetries, one has to look beyond inclusive $\pi^0$ production to the production of forward jets or direct photons. Present forward detector capabilities are not well matched to the complete reconstruction of forward jets, but do have sufficient acceptance for ``jet-like'' events.``Jet-like'' events are the clustered response of an electromagnetic calorimeter that is primarily sensitive to incident photons, electrons and positrons.

During the RHIC running in the year 2006, STAR with the Forward Pion Detector (FPD++) in place collected 6.8 pb$^{-1}$ of data with an average polarization of 60 \%. FPD++ was a modular detector prototype of the Forward Meson Spectrometer (FMS) that consisted of two detectors placed symmetrically with respect to the beam line at a distance of 7.4 m from the interaction point. Readout of the FPD++ was triggered when the sum of energies in the central module of the calorimeter used for $\pi^0$ measurements was larger than a threshold. This trigger minimizes the bias for ``jet-like'' events, making it appropriate to disentangling contributions to the forward transverse spin asymmetries. We report on the status of the analysis.
\end{abstract}

\section{Introduction}
The analyzing power ($A_N$) for a particle produced in a collision of transversely polarized protons is defined as the difference of spin-up and spin-down cross sections divided by their sum. The analyzing power $A_N$ is just one example of transverse single spin asymmetries, which is expected to be near zero in a leading-twist collinear perturbative QCD description of particle production. The measured cross sections for neutral pions ($\pi^0$) produced with large Feynman-x (2$p_L /\sqrt{s}$) and moderate $p_T$ in p+p collisions are found to be in agreement with next-to-leading order pQCD calculations at $\sqrt{s}=200\,$GeV \cite{Adams06, Arsene07}. ``Jet-like" structures are observed in two-particle correlations, involving a forward pion, as expected from pQCD. Precision measurements of the asymmetry as a function of $x_F$ and $p_T$ were reported, showing large $A_N$ at large $x_F$ \cite{Abelev08}. The measured $x_F$ dependence matches the Sivers effect \cite{Sivers90, Sivers91} expectations qualitatively, while the $p_T$ dependence at fixed $x_F$ is not consistent with expectations of p-QCD based calculations. 

Two concurrent models try to explain the observed asymmetries. The Sivers effect manifests itself as an asymmetry in the forward jet or gamma production while the Collins effect \cite{Collins93} manifests itself as an asymmetry in the forward jet fragmentation. To be able to distinguish between these effects, one has to look beyond inclusive $\pi$ events. Since the Collins effect is an azimuthal modulation of hadrons around the thrust axis of an outgoing quark, integrating over the azimuthal angle would cancel it, leaving only the Sivers effect. This is one possible path to distinguishing the two effects. A first attempt at separating the effects this way was done by looking at ``jet-like'' events at STAR with the FMS detector \cite{Poljak09}.``Jet-like" events are the clustered response of an electromagnetic calorimeter that is primarily sensitive to incident photons, electrons and positrons. The forward detector capabilities during the RHIC running in the year 2006 had sufficient acceptance for ``jet-like'' events. The data obtained were used to separate the Sivers/Collins contributions to the asymmetries observed for the $\pi$ events.

\section{Experimental setup}

The forward STAR calorimeters prior to RHIC running in the year 2006 measured the inclusive $\pi$ cross section as well as the single beam spin asymmetry for their inclusive production. These detectors, called Forward Pion Detector (FPD) were modular electromagnetic detectors that were positioned to make measurements at $<\eta>$ = 3.3, 3.7 or 4.0. In the year 2006 the FPD detectors placed on the West STAR platform were upgraded to a detector called the Forward Pion Detector ++ (FPD++), a prototype for the later Forward Meson Spectrometer (FMS).

\begin{figure}[h]
\begin{minipage}{16pc}
\includegraphics[width=16pc]{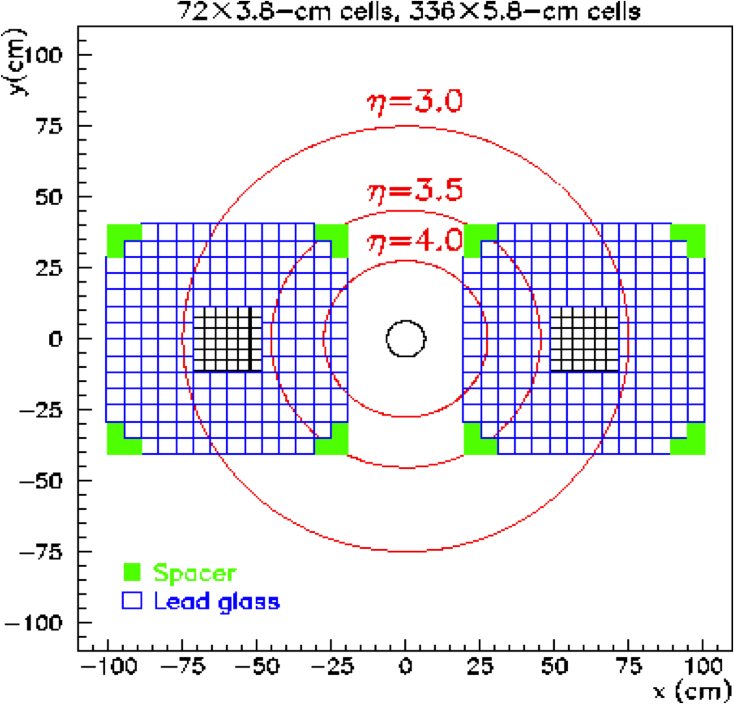}
\caption{\label{Scheme} The scheme of the FPD++.} 
\end{minipage}\hspace{2pc}
\begin{minipage}{19.9pc}
The FPD++ consisted of two modules placed symmetrically with respect to the beam line at a distance of 7.4 m from the interaction point, as drawn on Fig. \ref{Scheme}. The forward acceptance of STAR has been increased by $\approx$ 8 times by the FPD++. Each module was built from lead-glass detectors of two types: 180 cells of square cross section 5.8 cm on a side and 18 radiation lengths, and 36 cells of square cross section 3.8 cm on a side and 18 radiation lengths. The small cells were placed in the geometrical centers of the modules in square formations. The readout of the FPD++ was triggered when the sum of energies in a small cell module was larger than a threshold. The methodologies developed to calibrate the FPD have been extended to calibrate the FPD++. 
\end{minipage} 
\end{figure}

For the small cells, individual detector conversion gains are iteratively adjusted to give the Gaussian centroid at the $\pi$ mass in di-photon invariant mass distributions associated with each detector. The events used in the calibration are required to satisfy the hardware-level summed small cell energy threshold condition and have exactly two photons reconstructed within a fiducial volume of the calorimeter. The energies of the reconstructed photons are used to compute the invariant mass. The energy sharing, defined as $|E_1 - E_2|/(E_1+E_2)$, is required to be $<$ 0.7. 

In addition to adjustments of linear calibration factors for each of the small cells, energy-dependent and run-dependent corrections are made. The centroid of the $\pi^0$ peak in the mass distributions is found to increase with leading photon energy. The shower fit for the photons used for reconstruction does not completely account for transverse and longitudinal shower profiles at all energies and causes the reconstructed di-photon invariant mass to be energy dependent. To account for it, energy dependent corrections are applied to give the correct centroid for the $\pi^0$ peak in di-photon invariant mass distributions binned in leading photon energy. Reconstruction of full PYTHIA/GEANT simulations demonstrates that energy dependent corrections correctly reproduce the average energy of the $\pi^0$ when the centroid of the reconstructed di-photon invariant mass is at the measured value. Run-dependent corrections are made to account for small dependence of detector gain on count rate, proportional to luminosity decrease with time in store.

Given that the hardware trigger is only sensitive to the small cells, the large cells did not have many events that could be used to calibrate them in the same way as the small cells. To aid in calibration and to have a better understanding of the unpolarized result obtained in the data, full PYTHIA/GEANT simulations have been made with adequate statistics. The large cell calibration started by iteratively matching the slopes of the energy deposition histograms in data and simulations on a cell-by-cell basis. After obtaining the initial correction factors in this way, further calibration was done using the di-photon invariant mass spectra.

The stability of the calibration was checked with the help of a light-emitting diode (LED) monitoring system placed on the front face of each of the modules. The system can flash a number of LED signals at a given time so a response from the detector could be read out.

\section{Results}

To verify consistency with published work \cite{Abelev08}, two analyses were made. The $x_F$ dependence of $A_N$ for $p^{\uparrow} + p \rightarrow \pi +X$ from the small cells of the FPD++ was reanalyzed with a condition that the readout of large cells was ``live'' for the event. The small cells were read out by effectively deadtimeless ``flash ADC''. The large cells were read out by gated ADC, that introduced conversion and readout deadtime. The requirement that the large cell data was read out was required. Also, $\approx$ 20 \% of the statistics from the published work are excluded by quality assurance on the large cells. Published results did not include any information from the large cells of the FPD++. Reanalysis of the neutral pion $A_N$ is consistent with published results.

Properties of ``jet-like" events, obtained with the FPD++, were compared to those obtained previously with the FMS detector \cite{Poljak09}. After verifying consistency, a subset of data which contain both a reconstructed $\pi$ and a ``jet-like" object in a single event was extracted. For that subset, the Collins angle was calculated and the asymmetry dependence on the angle was found.

\begin{figure}[h]
\begin{minipage}{18pc}
\includegraphics[width=18pc]{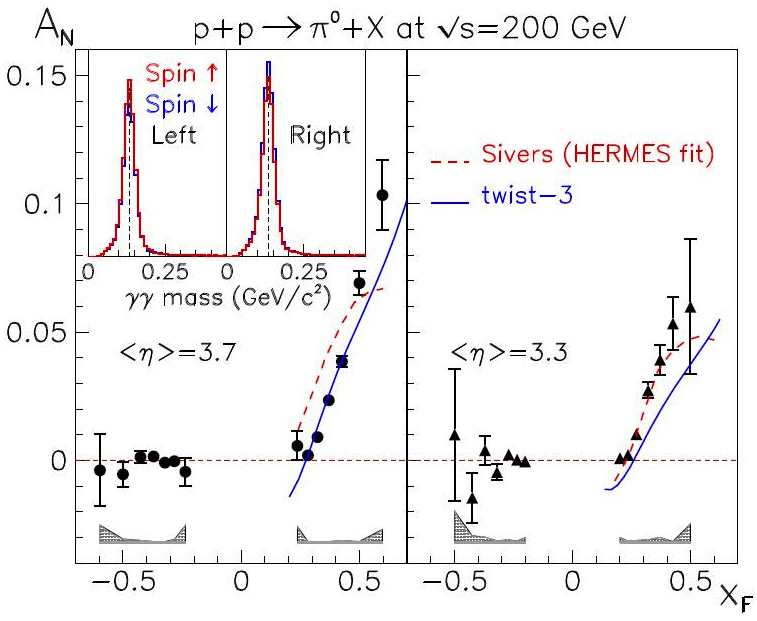}
\caption{\label{Run6res} The $x_F$ dependence of $A_N$, taken from \cite{Abelev08}. The left panel shows the results obtained with the East FPD detectors, the right with the West FPD detectors.} 
\end{minipage}\hspace{2pc}
\begin{minipage}{18pc}
\includegraphics[width=18pc]{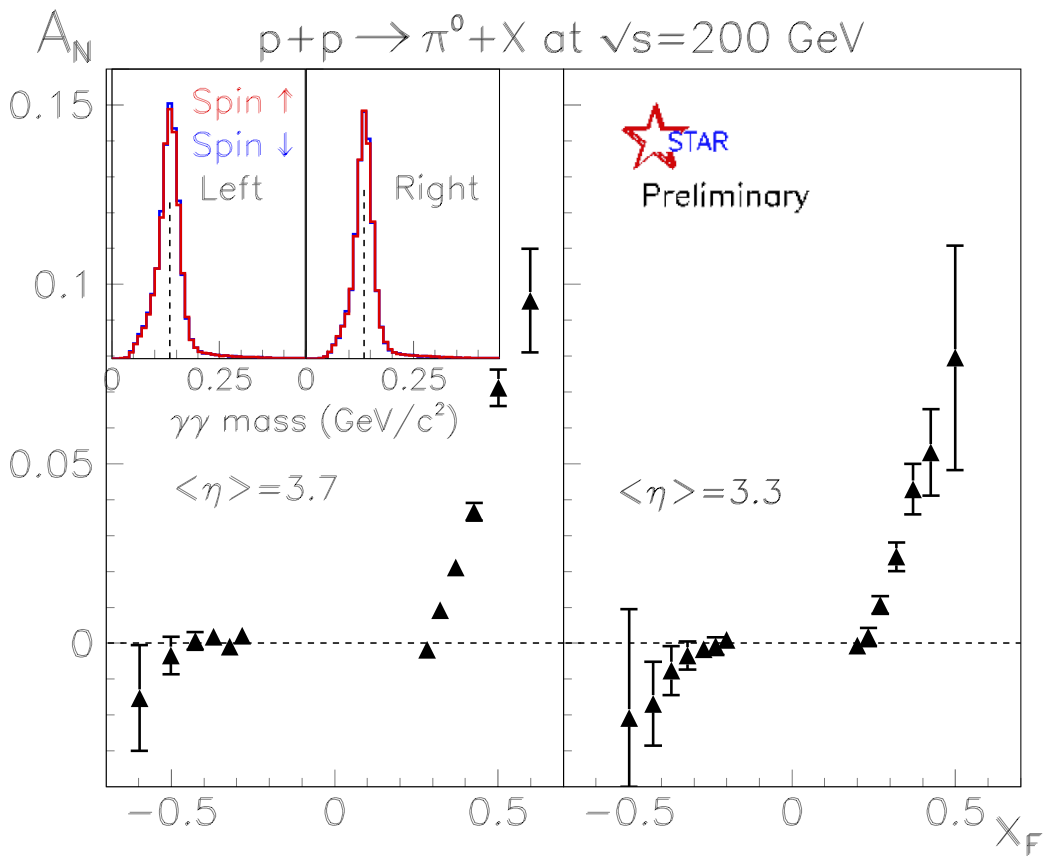}
\caption{\label{asyxflive}The $x_F$ dependence of $A_N$, as obtained with large cell ``live" readout condition. The missing lowest $x_F$ point in the left panel is due to decreased dataset.}
\end{minipage} 
\end{figure}

Figs. \ref{Run6res} and \ref{asyxflive} show results obtained from the East FPD detectors (left) and the West FPD/FPD++ detectors (right) to verify consistency. The published results include parts of data from RHIC running in the years 2003 and 2005, which is the reason for existence of the lowest $x_F$ point in the East detectors. The insets show examples of spin-sorted di-photon invariant masses, with the vertical line placed at the neutral pion rest mass. The error bars in the figures are statistical.

Comparison of ``jet-like'' event properties with results obtained with the FMS was also examined. ``Jet-like'' clusters are formed in an event by considering energy depositions $>$ 0.4 GeV in all cells of the FPD++. A cluster consists of N cells, where N is the maximal subset of cells found to be within a cone of radius 0.5 in $\eta - \phi$ space. The $x_F$ and $p_T$ for the cluster are given by the vector sum of momenta from each cell, assuming the energy deposition is from photons originating from the collision vertex. Clusters with at least 10 cells having cluster $p_T >$ 1.5 GeV/c and $x_F >$ 0.23 are required in the analysis. A further requirement that the cluster centroid is within the calorimeter volume by at least two large cell widths is also imposed. 


\begin{figure}[h]
\begin{minipage}{18pc}
\includegraphics[width=18pc]{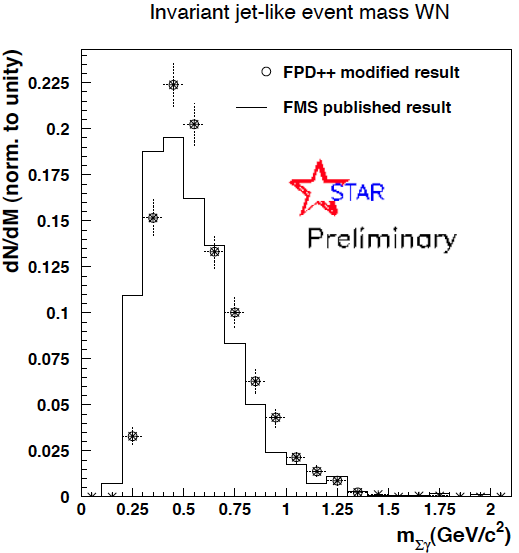}
\caption{\label{highmultmass}The comparison of the invariant mass spectrum of the ``jet-like'' events obtained with the FMS and the FPD++ with modifications.}
\end{minipage}\hspace{2pc}
\begin{minipage}{17.9pc}
The properties of the ``jet-like'' events that were looked at were the the energy deposition profile and the invariant mass distributions. The energy profiles show good agreement in the FPD++ and the FMS, as expected since it is a detector independent property. To compare the invariant mass spectra, two differences in the detectors had to be accounted for. The differences lie in the geometry of the detectors, which enters the event selection through a ``weighted multiplicity'' cut, and the hardware trigger of the detectors, which was a hightower trigger in the FMS and a summed module energy trigger in the FPD++. Both differences were modeled in the ``jet-like" event reconstruction algorithm to be able to compare the results. Good agreement for the jet-mass distributions from the FPD++ and the FMS is found, as seen in Fig. \ref{highmultmass}.
\end{minipage} 
\end{figure}


A subset of data from the FPD++ that contained both a reconstructed $\pi$ and a ``jet-like" object in a single event was extracted. The Collins angle $\gamma$ is defined as the azimuthal angle of the $\pi^0$ with respect to the reconstructed ``jet-like'' four-momentum. The angles $\gamma\approx 0$ are defined to be near the beam and the angles $\gamma\approx \pm \pi$ are defined to be far from it. ``Jet-like'' event properties suggest that the $\gamma$ distribution should peak near $\gamma =0$, since the $\pi^0$ carries most of the $p_T$ of the events. To verify that the direction of the reconstructed parton which fragments to the ``jet-like'' event is well reconstructed, an association analysis was done. The analysis is used to establish if the reconstructed ``jet-like'' event can be associated either with a hard-scattered parton or a parton from initial-state radiation. The result shows that the direction of the ``jet-like'' event is reconstructed well.

\begin{figure}[h]
\begin{minipage}{18pc}
\includegraphics[width=18pc]{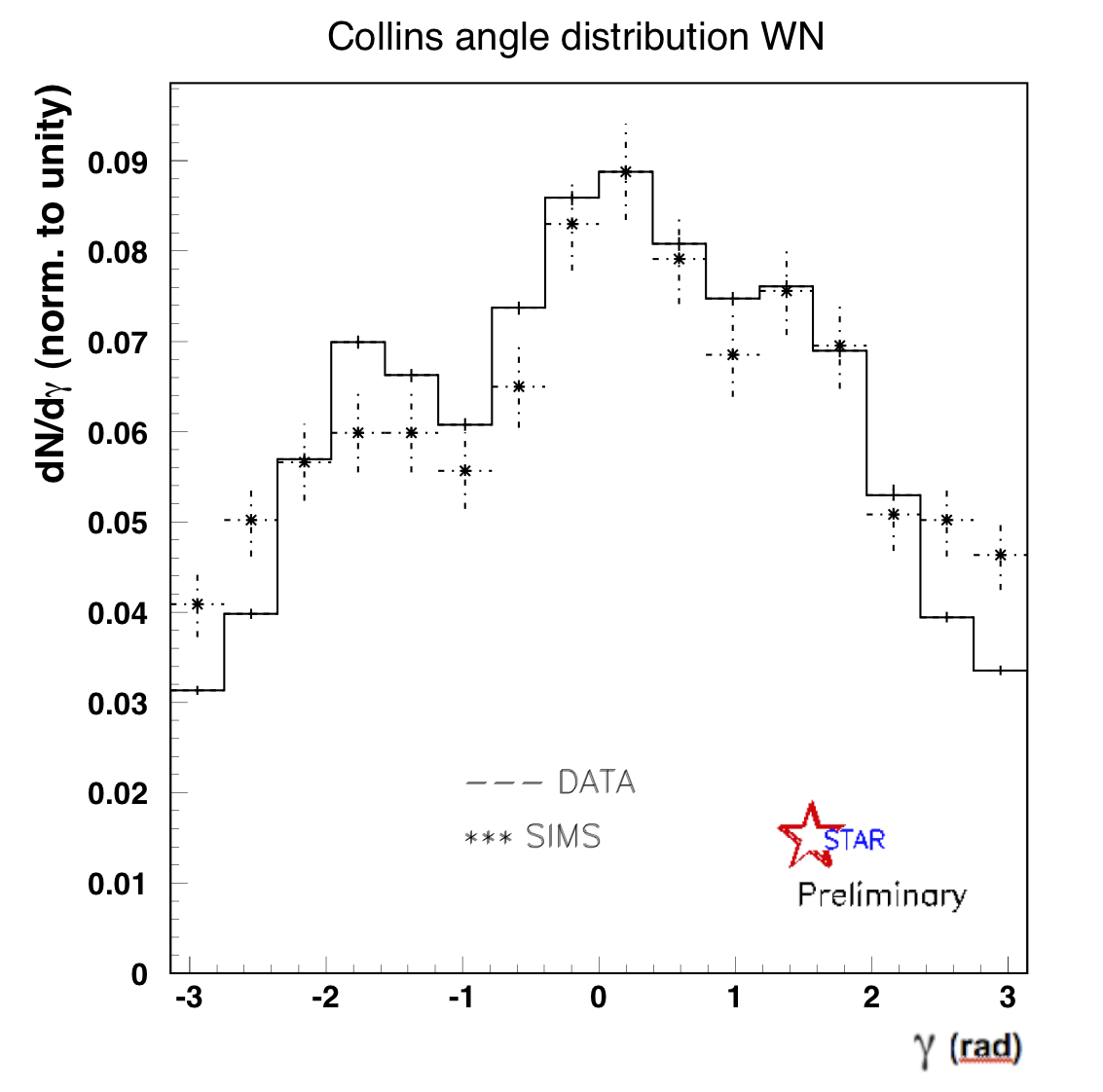}
\caption{\label{colldsim} The distribution of the Collins angle as obtained for one of the modules. The distribution is behaving as expected and the comparison of the data and simulations is reasonable.} 
\end{minipage}\hspace{2pc}
\begin{minipage}{18pc}
\includegraphics[width=18pc]{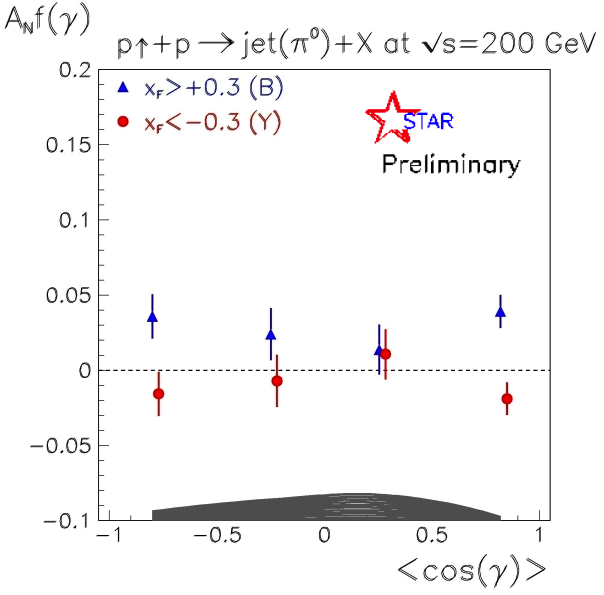}
\caption{\label{asycoscol}The dependence of the calculated asymmetry on $\cos(\gamma)$. The negative $x_F$ values are consistent with zero, while the positive $x_F$ values show a positive asymmetry, but no dependence on $\cos(\gamma)$. The lines represent statistical errors and the black area the systematic errors.}
\end{minipage} 
\end{figure}

The dependence of the asymmetry on the Collins angle was looked at. On the plot of $A_N$ vs. $\cos(\gamma)$ the Collins contribution would be proportional to the slope parameter. If the slope is consistent with zero and the asymmetry is larger than zero, one has isolated the Sivers effect, since the fragmentation is not involved, and one is left with just the initial state $k_T$. The results for the distribution of the Collins angle for a single module and the dependence of the calculated asymmetry on $\cos(\gamma)$ are given on Figs. \ref{colldsim} and \ref{asycoscol}.

The distribution of the Collins angle is behaving as expected and the comparison of the data and simulations is reasonable. The dependence of the calculated asymmetry on $\cos(\gamma)$ shows that the negative $x_F$ values are consistent with zero, while the positive $x_F$ values show a positive asymmetry, but no dependence on $\cos(\gamma)$. While there is a
small offset, the slope is consistent with zero. Hence, the Collins effect is not present in the production of forward neutral pions, isolating the Sivers effect.

\section{Conclusion}
The FPD++ was commissioned and operated in the RHIC running in the year 2006. The reconstruction and calibration procedures were successfully extended from the FPD and the FMS. The data show good agreement with the simulated sample of events. Both the inclusive pion $A_N(x_F)$ and the ``jet-like'' event properties are consistent with prior measurements. For the event sample that contains both a reconstructed neutral pion and a ``jet-like'' event, the Collins angle was calculated and the dependence of the asymmetry on the Collins angle was found. The final result shows that the Collins effect does not contribute to the measured asymmetries in the forward neutral pion production, isolating the Sivers contributions.

\section*{References}

\bibliography{Literature}

\providecommand{\newblock}{}
\begin{thebibliography}{1}
\expandafter\ifx\csname url\endcsname\relax
  \def\url#1{{\tt #1}}\fi
\expandafter\ifx\csname urlprefix\endcsname\relax\def\urlprefix{URL }\fi
\providecommand{\eprint}[2][]{\url{#2}}

\bibitem{Adams06}
Adams J 2006 {\em Phys. Rev. Lett.\/} {\bf 97} 152302

\bibitem{Arsene07}
Arsene I 2007 {\em Phys. Rev. Lett.\/} {\bf 98} 252001

\bibitem{Abelev08}
Abelev B~I 2008 {\em Phys. Rev. Lett.\/} {\bf 101} 222001

\bibitem{Sivers90}
Sivers D 1990 {\em Phys. Rev. D\/} {\bf 41} 83--90

\bibitem{Sivers91}
Sivers D 1991 {\em Phys. Rev. D\/} {\bf 43} 261--3

\bibitem{Collins93}
Collins J~C 1993 {\em Nucl. Phys. B\/} {\bf 396} 161--82

\bibitem{Poljak09}
Poljak N 2009 {\em SPIN08 Conf. Proceed.\/}  521--4

\end{thebibliography}

\end{document}